\documentclass[preprint]{revtex4}
\usepackage{amsfonts}
\usepackage{color,graphicx,amssymb,amsbsy,amsmath}

\begin{document}

\title{Effects of population mixing on the spread of SIR epidemics}
\author{Henryk Fuk\'s}
\email{hfuks@brocku.ca}
\affiliation{Department of Mathematics, Brock University\\
St. Catharines, Ontario L2S 3A1, Canada }
\author{Anna T. Lawniczak}
\email{alawnicz@uoguelph.ca}
\affiliation{Department of Mathematics and Statistics\\
University of Guelph\\
Guelph, Ontario N1G 2W1, Canada}
\author{Ryan Duchesne}
\affiliation{Department of Mathematics, Brock University \\
St. Catharines, Ontario L2S 3A1, Canada }
\date{\today }

\begin{abstract}
\noindent \emph{Abstract: - }We study dynamics of spread of epidemics of SIR
type in a realistic spatially-explicit geographical region, Southern and
Central Ontario, using census data obtained from Statistics Canada, and
examine the role of population mixing in epidemic processes. Our model
incorporates the random nature of disease transmission, the discreteness and
heterogeneity of distribution of host population.We find that introduction
of a long-range interaction destroys spatial correlations very easily if
neighbourhood sizes are homogeneous. For inhomogeneous neighbourhoods, very
strong long-range coupling is required to achieve a similar effect. Our work
applies to the spread of influenza during a single season.
\end{abstract}

\maketitle




\section{Introduction}

Studies of some epidemics, for example, the spread of the Black Death in
Europe from 1347--1350 \cite{Langer1964}, the past influenza pandemics \cite%
{Rvachev95}, spread of fox rabies in Europe, or spread of rabies among
raccoons in eastern United States and Canada \cite{Murray2003b}, indicate
that host and infective interactions and spatial distributions of their
populations should play an important role in the dynamics of spread of many
infectious diseases.

Until recently most mathematical models of spread of epidemics have
described interactions of large number of individuals in aggregate form and
often these models have neglected aspects of spatial distribution of
populations, importance of which have been addressed in \cite{durret94}.
Adopting methodologies like cellular automata, coupled map lattices, lattice
gas cellular automata or agent based simulations, new classes of models have
been proposed and studied \cite{paper15,paper16,benyo2000,paper24,Watts2004,Barrett2005}, to incorporate
with various levels of abstraction and details: direct interactions among
individuals; spatial distribution of population types (i.e., infective,
susceptible, removed); individuals' movement; effects of social networks on
spread of epidemics.

The goal of our work is to study the effects of population interactions and
mixing on the spatio-temporal dynamics of spread of epidemics of SIR
(susceptible-infected-removed) type in a realistic population distribution.
For the purpose of our study we developed a fully discrete
individually-based simulation model that incorporates the random nature of
disease transmission. The key feature of this model is the fact that for
each individual the set of all individuals with whom he/she interacts may
change with time. This results in time varying small world network structure.

As a case study we considerder census data obtained from Statistic Canada 
\cite{statcan1,statcan2} for Southern and Central Ontario. The data
set specifies population of small areas composed of one or more neighbouring
street blocks, called \textquotedblleft dissemination
areas\textquotedblright . Using these data, we study the effects of two
types of interactions among individuals on the spread of epidemics. The
first type of interaction is the one among individuals located only in
adjacent dissemination areas. The second type of interaction is the one
among individuals who in addition to being in contact with members of their
own and adjacent dissemination areas may also be in contact with individuals
located in remote, non adjacent, dissemination areas. This last case can be
seen as a case of "short-cuts" among multiple far away dissemination areas.
We investigate spatial correlations in our model and how they can be
destroyed by the "short-cuts" in population contacts. Additionally, we
derive a mean field description of our individually based simulation model
and compare the results of the two models.

The presented work is continuation and expansion of our work in \cite%
{paper15,paper16,paper24} and contributes to better understanding of spread
of epidemics of SIR type, including influenza.

\section{SIR\ epidemic model description}

In order to study how population interactions ("population mixing") affects
the spread of epidemics we construct an individually based model in which
each individual is represented by a particle, as in our earlier work \cite%
{paper15,paper16}. Models of this type take various forms, ranging from
stochastic interacting particle systems \cite{durret94} to models based on
cellular automata or coupled map lattices~\cite%
{schon93,bc93,duryea99,benyo2000}.

In our model, we consider a set of $N$ individuals, labelled with
consecutive integers $1,2\ldots ,N$. This set of labels is denoted by $%
\mathcal{L}$. We assume that each individual, at any given time can be in
one of three, mutually exclusive, distinct states: susceptible (S), infected
(I) or removed (R). An individual can change state only in two ways: a
susceptible individual who comes in direct contact with an infected
individual can become infected with probability $p$; an infected individual
can become removed with probability $q$. The precise description of the
model is as follows.

The state of the $i$-th individual at the time step $k$ is described by a
Boolean vector variable $\boldsymbol{\eta }(i,k)=\langle \eta
_{S}(i,k),\eta _{I}(i,k),\eta _{R}(i,k)\rangle $, where $\eta _{\tau
}(i,k)=1 $ if the $i$-th individual is in the state $\tau $, where $\tau \in
\{S,I,R\} $, and $\eta _{\tau }(i,k)=0$ otherwise. Thus,
\begin{equation*}
\eta _{S}(i,k) \oplus \eta _{I}(i,k) \oplus \eta _{R}(i,k)=1.
\end{equation*}%
We assume that $i=1,2,\ldots ,N$ and $k\in \mathbb{N}$, that is, the time is
discrete. Hence, the only allowed values of the vector $\boldsymbol{\eta }%
(i,k)$ are
\begin{equation*}
\boldsymbol{\eta }(i,k)=\left\{ 
\begin{array}{c}
\langle 1,0,0\rangle ,\text{ \ \ \ \ \ \ \ if the }i\text{\ -th individual's
state }\in S \\ 
\langle 0,1,0\rangle ,\text{ \ \ \ \ \ \ \ \ \ if the }i\text{\ -th
individual's state }\in I \\ 
\langle 0,0,1\rangle ,\text{ \ \ \ \ \ \ \ \ if the }i\text{\ -th
individual's state }\in R%
\end{array}%
\right. .
\end{equation*}
No other values of $\boldsymbol{\eta }(i,k)$ are possible in SIR epidemic
model. In SIR model an individual can only be at one state at any give time
and transitions occur only from susceptible to infected and from infected to
removed. A removed individual does not become susceptible or infected again
in SIR model. Therefore, SIR model is suitable for studying spread of
influenza in the same season because the same type of influenza virus can
infect an individual only once and once the individual is recovered from the
flu it becomes immune to this type of virus.

We further assume that at time step $k$ the $i$-th individual can interact
with individuals from a subset of $\mathcal{L}$, to be denoted by $C(i,k)$.
Using this notation, after one time iteration the $\eta _{\tau }(i,k+1)$
becomes 
\begin{eqnarray}
\eta _{S}(i,k+1) &=&\eta _{S}(i,k)\prod_{j\in C(i,k)}(1-X_{i,j,k}\eta
_{I}(j,k)),  \label{sirdyn1} \\
\eta _{I}(i,k+1) &=&\eta _{S}(i,k)\left( 1-\prod_{j\in
C(i,k)}(1-X_{i,j,k}\eta _{I}(j,k))\right) +\eta _{I}(i,k)(1-Y_{i,k}), \label{sirdyn2} \\
\eta _{R}(i,k+1) &=&\eta _{R}(i,k)+\eta _{I}(i,k)Y_{i},\label{sirdyn3}
\end{eqnarray}%
where $X=\{X_{i,j,k}:i,j=1,...,N$ and $k=1,2,....\}$ is a sequence of iid
Boolean random variables such that $Pr(X_{i,j,k}=1)=p$, $Pr(X_{i,j,k}=0)=1-p$%
, and $Y=\{Y_{i,k}:i=1,...,N,$ and $k=1,2,...\}$ is a sequence$\ $of iid
Boolean variables such that $Pr(Y_{i}=1)=q$, $Pr(Y_{i}=0)=1-q$. We assume
that the sequences $X$ and $Y$ of random variables are independent of each
other and of the random variables $\eta _{\tau }(i,k).$

Observe that, if the $i$-th individual interacts with an infected $j$-th
individual at time step $k$ and $X_{i,j,k}=1,$ then the infection is
transmitted from the $j$-th individual to the $i$-th individual at this time
step. Thus, if some product $X_{i,j,k}\eta _{I}(j,k)$ takes the value 1,
then $\eta _{S}(i,k+1)=0$, meaning that the $i$-th individual has changed
its state from susceptible to infected.

The key feature of this model is the set $C(i,k)$, representing all
individuals with whom the $i$-th individual may have interacted at time step 
$k$. In a large human population, it is almost impossible to know $C(i,k)$
for each individual, so we make some simplifying assumptions. First of all,
it is clear that the spatial distribution of individuals must be reflected
in the structure of $C(i,k)$. We have decided to use realistic population
distribution for Southern and Central Ontario using census data obtained
from Statistic Canada \cite{statcan1,statcan2}. The selected region is
mostly surrounded by waters of Great Lakes, forming natural boundary
conditions. The data set specifies population of so called \textquotedblleft
dissemination areas\textquotedblright\ , that is, small areas composed of
one or more neighbouring street blocks. We had access to longitude and
latitude data with accuracy of roughly $0.01^{\circ }$, hence some
dissemination areas in densely populated regions have the same geographical
coordinates. We combined these dissemination areas into larger units, to be
called \textquotedblleft modified dissemination areas\textquotedblright\
(MDA).

We now define the set $C(i,k)$ using the concept of MDAs. This set is
characterized by two positive integers $n_{c}$ and $n_{f}$. Let us label all
MDAs in the region we are considering by integers $m=1,2,\ldots ,M$, where
in our case $M=5069$. For the $i$-th individual belonging to the $m$-th MDA,
the set $C(i,k)$ consists of all individuals belonging to the $m$-th MDA
plus all individuals belonging to the $n_{c}$ MDAs nearest to the $m$-th MDA
and the $n_{f}$ MDAs randomly selected among all remaining MDAs. While the
\textquotedblleft close neighbours\textquotedblright , that is, the $n_{c}$
nearest MDAs, will not change with time, the \textquotedblleft far
neighbours\textquotedblright , that is, the $n_{f}$ randomly selected MDAs,
will be randomly reselected at each time step.

\section{Derivation of mean field equations}

The model described in the previous section involves strong spatial coupling
between individuals. Before we describe consequences of this fact, we first
construct a set of equations which approximate dynamics of the model under
the assumption of \textquotedblleft perfect mixing\textquotedblright , in
other words, neglecting the spatial coupling.

The state of the system described by eq. (\ref{sirdyn1}--\ref{sirdyn3}) at
time step $k$ is determined by the states of all individuals and is
described by the Boolean random field $\boldsymbol{\eta }(k)=\{\boldsymbol{%
\eta }(i,k):i=0,\ldots ,N\}$. Under the assumptions of our model, the
Boolean field $\{\boldsymbol{\eta }(k):i=0,1,2\ldots \}$ is a Markov
stochastic process.

By taking the expectation $E_{\boldsymbol{\eta }(0)}$ of this Markov
stochastic process when the initial configuration is $\boldsymbol{\eta }(0)$
we get the probabilities of the $i$-th individual being susceptible, or
infected, or removed at time $k$, that is, $\rho _{\tau }(i,k)=E_{%
\boldsymbol{\eta }(0)}\left[ \eta _{\tau }(i,k)\right] $ for $\tau \in
\{S,I,R\}$.

Since the sequences of random variables $X$ and $Y$ are independent of each
other and of the sequences of the random variables $\eta _{\tau }(i,k)$,
assuming additionally independence of random variables $\eta _{\tau }(i,k)$,
the expected value of a product of these variables is equal to the product
of expected values. Under these mean field assumptions, taking expected
values of both sides of equations (\ref{sirdyn1}--\ref{sirdyn3}) we obtain 
\begin{eqnarray}
\rho _{S}(i,k+1) &=&\rho _{S}(i,k)\prod_{j\in C(i,k)}(1-p\rho _{I}(j,k)),
\label{sirmf3} \\
\rho _{I}(i,k+1) &=&\rho _{S}(i,k)\Big(1-\prod_{j\in C(i,k)}(1-p\rho
_{I}(j,k))\Big)+\rho _{I}(i,k)(1-q), \\
\rho _{R}(i,k+1) &=&\rho _{R}(i,k)+\rho _{I}(i,k)q.
\end{eqnarray}%
Since mean field approximations neglect spatial correlations, we further
assume that $\rho _{\tau }(i,k)$ is independent of $i$, that is $\rho _{\tau
}(i,k)=\rho _{\tau }(k)$. Even though sets $C(i,k)$ have different number of
elements for different $i$ and $k$, for the purpose of this approximate
derivation we assume that they all have the same number of elements $%
(1+n_{c}+n_{f})D$, where $D$ is the average MDA population. All these
assumptions lead to 
\begin{align}
\rho _{S}(k+1)& =\rho _{S}(k)(1-p\rho _{I}(k))^{(1+n_{c}+n_{f})D},
\label{mf1} \\
\rho _{I}(k+1)& =\rho _{I}(k)+\rho _{S}(k)-\rho _{S}(k)(1-p\rho
_{I}(k))^{(1+n_{c}+n_{f})D}-q\rho _{I}(k), \label{mf2}\\
\rho _{R}(k+1)& =\rho _{R}(k)+q\rho _{I}(k).\label{mf3}
\end{align}%
The third equation in the above set is obviously redundant, since $\rho
_{S}(k)+\rho _{I}(k)+\rho _{R}(k)=1$.

Similarly to the classical Kermack-McKendrick model, mean field equations (%
\ref{mf1})-(\ref{mf3}) exhibit a threshold phenomenon. Depending on the
choice of parameters, we can have $\rho _{I}(k)<\rho _{I}(0)$ for all $k$,
meaning that the infection is not growing and eventually it will die out
because in our model no new individuals are being born or arrive from
outside the area under consideration during the time of the epidemic.
Alternatively, we can have $\rho _{I}(k)>\rho _{I}(0)$ for some $k$, meaning
that the epidemic is spreading. The intermediate scenario of constant $\rho
_{I}(k)$ will occur when $\rho _{I}(k)=\rho _{I}(0)$, that is, when 
\begin{equation}
\rho _{S}(0)-\rho _{S}(0)(1-p\rho _{I}(0))^{(1+n_{c}+n_{f})D}-q\rho
_{I}(0)=0.  \label{trecond}
\end{equation}%
Assuming that initially the entire population consists only of susceptible
and infective individuals, that is, there are no individuals in the removed
group at $k=0,$we have $\rho _{S}(0)=1-\rho _{I}(0)$. Furthermore, if $%
(1+n_{c}+n_{f})D$ is large, we can assume $(1-p\rho
_{I}(0))^{(1+n_{c}+n_{f})D}\approx 1-p(1+n_{c}+n_{f})D\rho _{I}(0)$. Solving
eq. (\ref{trecond}) for $q$ under these assumptions we obtain 
\begin{equation}
q=\Big(1-\rho _{I}(0)\Big)(1+n_{c}+n_{f})Dp.  \label{separatMF}
\end{equation}%
Thus, assuming the mean field approximation the epidemic can occur only if $%
q<\Big(1-\rho _{I}(0)\Big)(1+n_{c}+n_{f})Dp$.

\section{Spatio-temporal dynamics of SIR epidemic model}

\begin{figure}[tbp]
\centering
\includegraphics[scale=0.7]{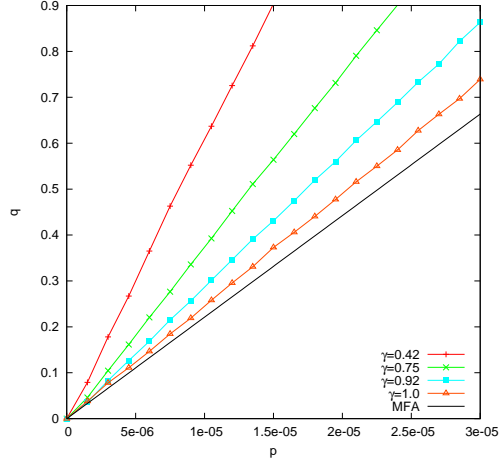}
\caption{Graphs of critical lines for $\protect\gamma =0.42,0.75,0.92$, and $%
1.0$. The first line from the bottom represents mean field approximation.}
\label{phasetran1}
\end{figure}

\begin{figure}[t]
\centering
\includegraphics[scale=0.43]{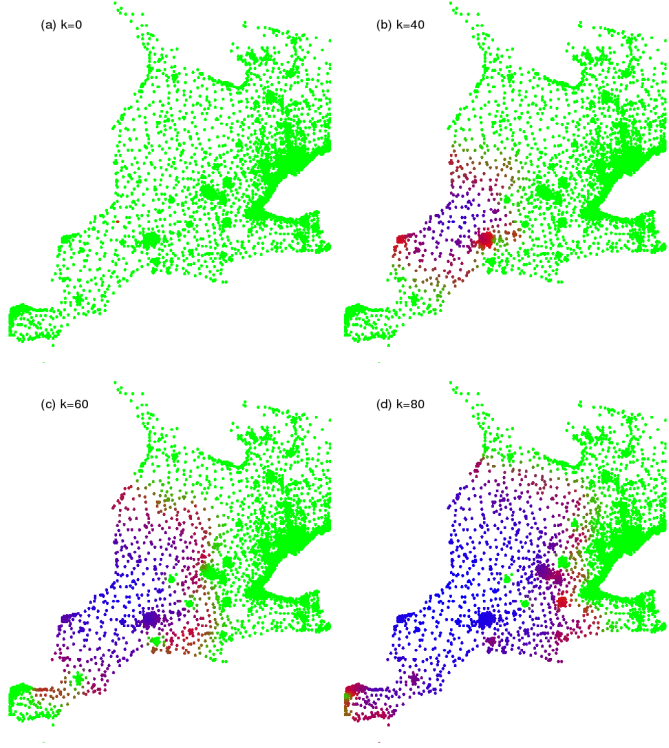}
\caption{Example of a propagating epidemic front for $\protect\gamma =0$, $%
p=0.00005$, $q=0.05$, with (a) $k=0$, (b) $k=40$, (c) $k=60$ and (d) $k=80$.
The initial outbreak is located in an area with low population density.
Modified dissemination areas are represented by pixels colored according to
density of individuals of a given type, such that the red component
represents density of infected, green density of susceptibles, and blue
density of removed individuals.}
\label{fig:frontlow}
\end{figure}

The mean-field equations derived in the previous section depend only on the
sum of $n_{c}$ and $n_{f}$. This means, for example, that the model with $%
n_{c}=12$, $n_{f}=0$ and the model with $n_{c}=11$, $n_{f}=1$ will have the
same mean field equations. However, the actual dynamics in these two cases
are very different, see Figure \ref{fig:frontlow} and Figure \ref%
{fig:frontdest}. Depending on the relative size of $n_{f}$ and $n_{c}$, the
epidemic may propagate or die out, as the following analysis shows. In order
to make the subsequent analysis more convenient, we introduce parameter $%
\gamma $, defined as 
\begin{equation}
\gamma =\frac{n_{f}}{n_{c}+n_{f}}.  \label{gamma}
\end{equation}%
Let $N_{\tau }(k)$ be the expected value of the total number of individuals
belonging to class $\tau \in \{S,I,R\}$, that is, 
\begin{equation*}
N_{\tau }(k)=E_{\boldsymbol{\eta }(0)}\left( \sum_{i=1}^{N}\eta _{\tau
}(i,k))\right) =\sum_{i=1}^{N}\rho _{\tau }(i,k).
\end{equation*}%
We say that an epidemic occurs if there exists $k>0$ such that $%
N_{I}(k)>N_{I}(0)$. For fixed $p$, $n_{f}$ and $n_{c}$, there exists a
threshold value of $q$ to be denoted by $q_{c}$, such that for each $q<q_{c}$
an epidemic occurs, and for $q>q_{c}$ it does not occur. Obviously $q_{c}$
depends on $p$, and this is illustrated in Figure~\ref{phasetran1}, which
shows graphs of $q_{c}$ as a function of $p$ for several different values of 
$\gamma $, where $n_{f}+n_{c}=12$. The graphs were obtained numerically by
direct computer simulations of the model. The condition $n_{f}+n_{c}=12$
means that the size of the neighbourhood is kept constant, but the
proportion of \textquotedblleft far neighbours\textquotedblright\
(represented by $\gamma $) varies. 
Figure~\ref{phasetran1} also shows the mean-field line given by eq. (\ref%
{separatMF}).

\begin{figure}[t]
\centering
\includegraphics[scale=0.43]{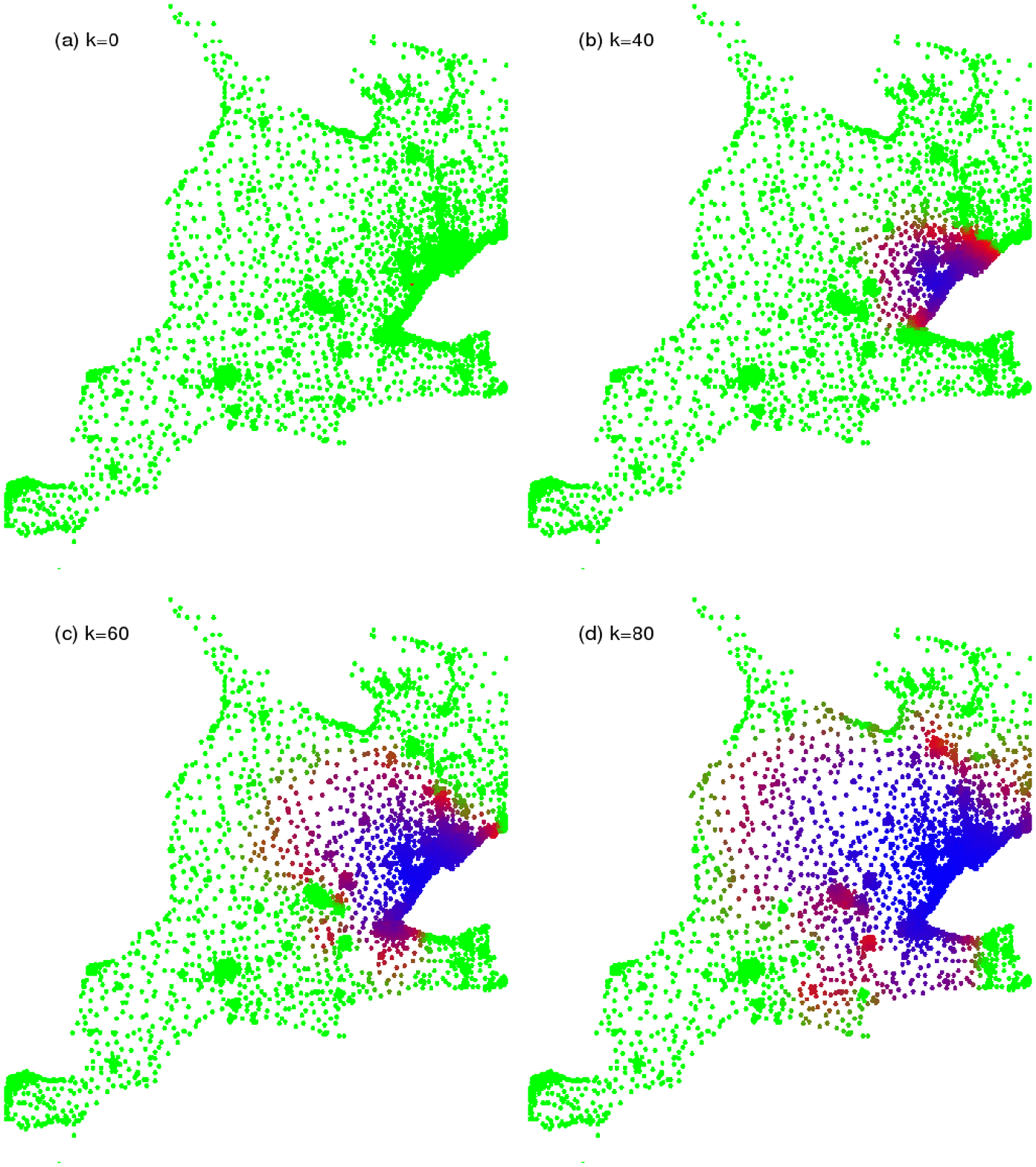}
\caption{Example of a propagating epidemic front parameters identical as in
Figure 2, except that the initial outbreak is now located in an area with
high population density. Color coding like in the previous figure.}
\label{fig:fronthigh}
\end{figure}

We observe that the parameter $\gamma $ controls dynamics of the epidemic
process in a significant way, shifting the critical line up or down. When $%
\gamma =0$, that is, when there are no interactions with \textquotedblleft
far neighbours\textquotedblright , the epidemic process has a strictly local
nature, and we can observe well defined epidemic fronts propagating in
space, regardless at which MDA the epidemic starts at $k=0$. This is
illustrated in Figure \ref{fig:frontlow}, where the epidemic starts at a
single centrally located MDA with low population density (Figure \ref%
{fig:frontlow}a) and on Figure \ref{fig:fronthigh}, where the epidemic
starts in a MDA with high population density (Figure \ref{fig:fronthigh}a).
The simulations were done for the same parameters in both cases except for
the different locations of the onsets of epidemics. The figures display MDAs
that are represented by pixels colored according to the density of
individuals of a given type. The red component of the color represents
density of infected individuals, green density of susceptibles, and blue
density of removed individuals. By density we mean the number of individuals
of a given type divided by the size of the population of the MDA. 
The epidemic waves propagating outwards can be clearly seen on Figure \ref%
{fig:frontlow} and Figure \ref{fig:fronthigh}, in the successive snapshots
(b), (c) and (d). The fronts are mostly red. This means that the bulk of
infected individuals is located at the fronts. After these individuals
gradually recover the centers become blue.

Let us now consider slightly modified parameters, taking $\gamma =\frac{1}{12%
}$. This means that we now replace one \textquotedblleft
close\textquotedblright\ MDA by one \textquotedblleft far\textquotedblright\
MDA. This does not seem to be a significant change, yet the effect of this
change is truly noticeable. 
\begin{figure}[t]
\centering
\includegraphics[scale=0.43]{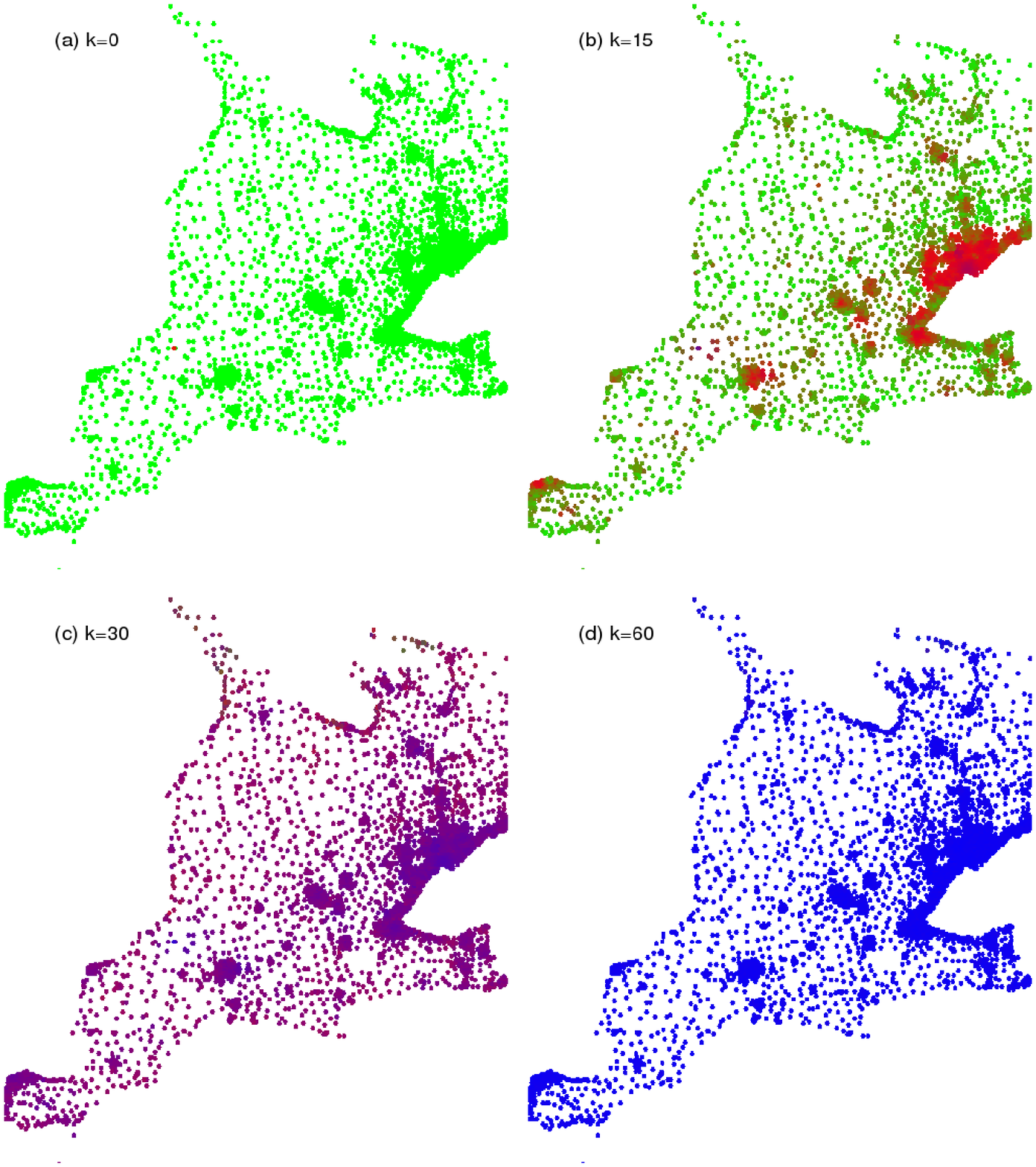}
\caption{Development of the epidemic for $\protect\gamma =\frac{1}{12}$, $%
p=0.00005$, $q=0.05$, with (a) $k=0$, (b) $k=15$, (c) $k=30$ and (d) $k=60$.
Colour coding is the same as in the previous figure.}
\label{fig:frontdest}
\end{figure}
As we can see in Figure~\ref{fig:frontdest}, the epidemic propagates much
faster, and there are no visible fronts. The disease quickly spreads over
the entire region and large metropolitan areas become red in a short time,
as shown in Figure~\ref{fig:frontdest}(b). This suggests that infected
individuals are more likely to be found in densely populated regions, and
their distribution is dictated by the population distribution -- unlike in
Figure~\ref{fig:frontlow} or Figure~\ref{fig:fronthigh}, where infected
individuals are to be found mainly at the propagating front.

\section{Spatial correlations of SIR epidemic model}

In order to quantify the observations of the previous section, we use a
spatial correlation function for densities of infected individuals defined
as 
\begin{equation*}
h(r,k)=\left\langle \eta _{I}(i,k)\eta _{I}(j,k)\right\rangle _{r\leq
d(i,j)\leq r+\Delta r},
\end{equation*}%
where $d(i,j)$ is the distance between $i$-th and $j$-th individual, and $%
<\cdot >$ represents averaging over all pairs $i$, $j$ satisfying condition $%
r\leq d(i,j)\leq r+\Delta r$. In the following considerations we take $%
\Delta r=1\,\mathrm{km}$. The distance between two individuals is defined as
the distance between MDAs to which they belong.

\begin{figure}[tbp]
\begin{center}
\includegraphics[scale=0.9]{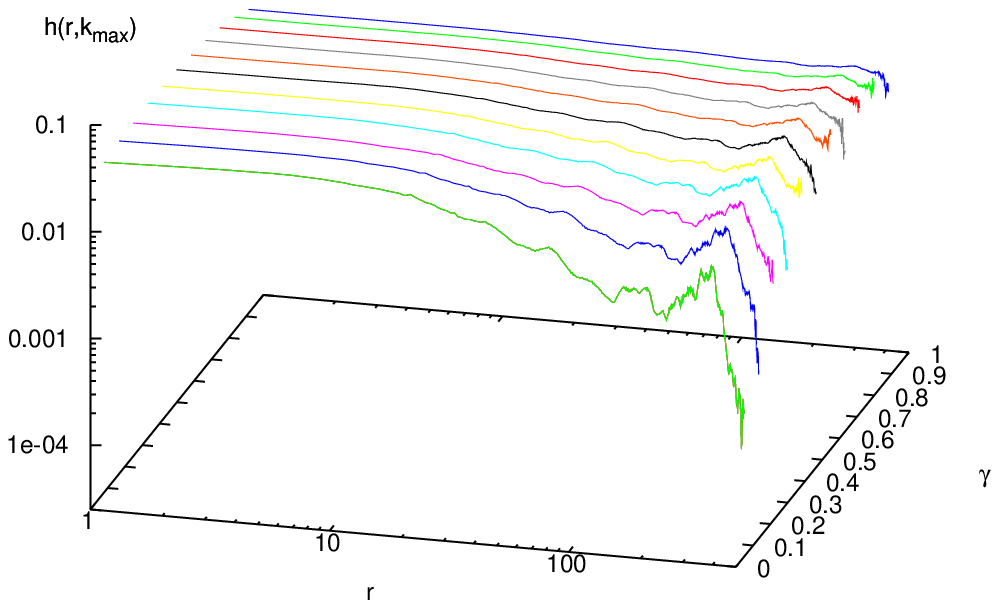}
\end{center}
\caption{Graphs of the correlation function $h(r,k_{max})$ for different
values of $\protect\gamma $, where $p=0.000015$, $q=0.2$, and $%
n_{c}+n_{f}=12 $.}
\label{cor3d}
\end{figure}

\begin{figure}[tbp]
\begin{center}
\includegraphics[scale=0.7]{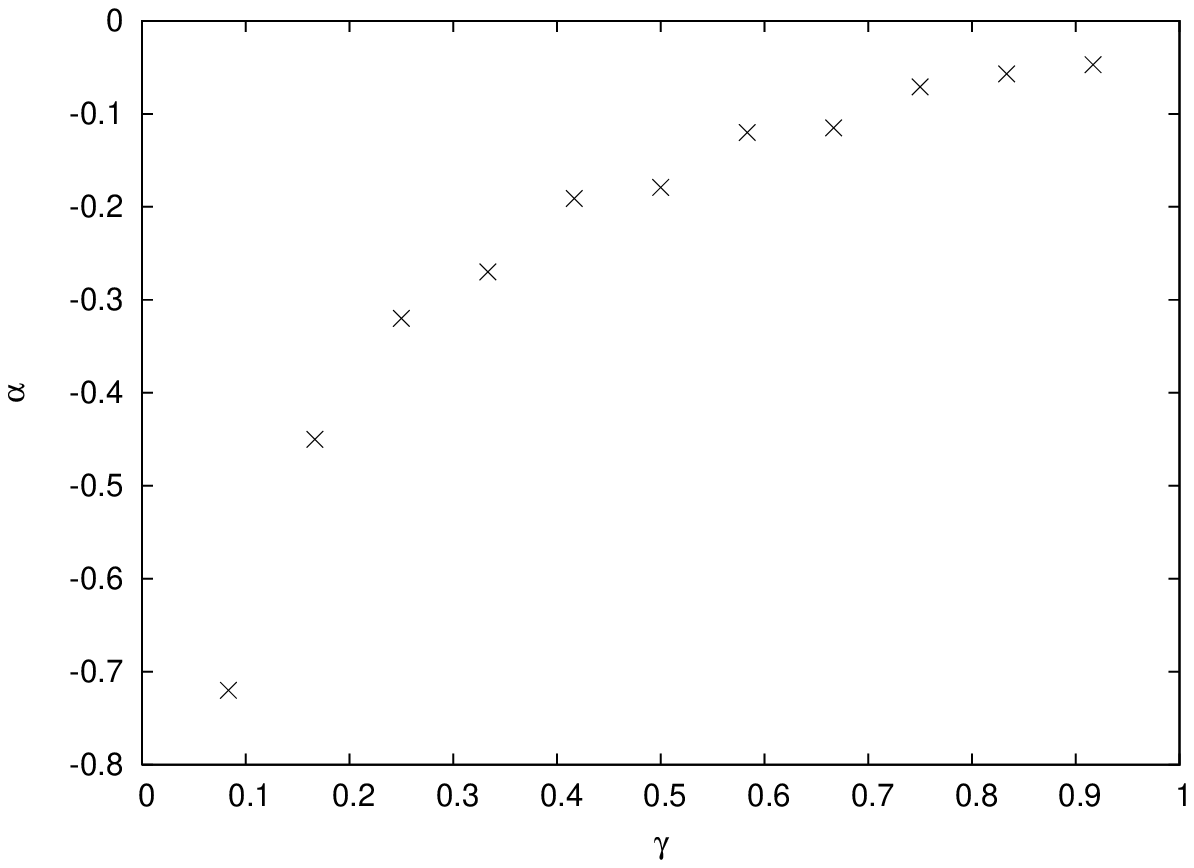}
\end{center}
\caption{Values of the exponent $\protect\alpha $ for different values of
the parameter $\protect\gamma $. The exponent has been obtained by fitting $%
h(r,k_{max})=Cr^\alpha $ to simulation data.}
\label{alpha}
\end{figure}

Consider now a specific example of the epidemic process described by eq. (%
\ref{sirdyn1}-\ref{sirdyn3}), where $p=0.000015$, $q=0.2$, and $%
n_{c}+n_{f}=12$. For this choice of parameters epidemics always occur as
long as $\gamma >0$. Figure \ref{cor3d} shows graphs of the correlation
functions $h(r,k_{max})$ at the peak of each epidemic, so that $k_{max}$ is
the time step at which the number of infected individuals achieves its
maximum value. 
An interesting phenomenon can be observed in the figure under consideration:
while the increase of the proportion of \textquotedblleft
far\textquotedblright\ neighbours does destroy spatial correlations, one
needs very high proportion of \textquotedblleft far\textquotedblright
neighbours to make the correlation curve completely flat. In \cite%
{Bonabeau1998} it is reported that for influenza epidemics $h(r,k_{max})\sim
r^{0.04\pm 0.03}$. If we fit $h(r,k_{max})=Cr^{\alpha }$ curve to the
correlation data shown in Figure \ref{cor3d}, we obtain values of the
exponent $\alpha $ as shown in Figure \ref{alpha}. In order to obtain $%
\alpha $ of comparably small magnitude as reported in \cite{Bonabeau1998},
one would have to take $\gamma $ equal to at least $0.83$, meaning that vast
majority of neighbours would have to be \textquotedblleft far
neighbours\textquotedblright . 
In reality, this would require that the vast majority of all individuals one
interacted with were not his/her neighbours, coworkers, etc., but
individuals from randomly selected and possibly remote geographical regions.
This is clearly at odds with our intuition regarding social interactions,
especially outside large metropolitan areas. This prompted us to investigate
further and to find out what is responsible for this effect.

Upon closer examination of spatial patterns generated in simulations of our
individually-based model, we reach the conclusion that the inhomogeneity of
population sizes in neighbourhoods $C(i,k)$ makes spatial correlations so
persistent. Since different MDAs have different population sizes, we expect
that some individuals will have larger neighbourhood populations than
others, and as a result they will be more likely to get infected, even if
the proportion of infected individuals is the same in all MDAs. This will
build up clusters of infected individuals around populous MDAs.

To test if this is indeed the factor responsible for strong spatial
correlations in our model, we replaced all MDA population sizes with a
constant population size $D$, that is, the average MDA population size. As
expected, graphs of the correlation functions obtained in this case were are
all essentially flat, with the exponent $\alpha $ close to zero even in the
case of $n_{f}=1$, when we obtained $\alpha =0.023\pm 0.002$.

\section{Conclusions}

We conclude that spatial correlations are difficult to destroy if
neighbourhood sizes are inhomogeneous. Very significant amount of long-range
interactions (i.e., very strong mixings) is required to obtain flat
correlations curves. However, for homogeneous neighbourhood sizes, even
relatively small long-range interaction immediately forces the process into
the perfect-mixing regime, resulting in the lack of spatial correlations.
The model that we developed can be applied to other realistic population
distributions and geographical regions.

\paragraph*{Acknowledgement}

Henryk Fuk\'s and Anna T. Lawniczak acknowledge partial financial support from the Natural Science
and Engineering Research Council (NSERC) of Canada.


\begin{thebibliography}{16}
\expandafter\ifx\csname natexlab\endcsname\relax\def\natexlab#1{#1}\fi
\expandafter\ifx\csname bibnamefont\endcsname\relax
  \def\bibnamefont#1{#1}\fi
\expandafter\ifx\csname bibfnamefont\endcsname\relax
  \def\bibfnamefont#1{#1}\fi
\expandafter\ifx\csname citenamefont\endcsname\relax
  \def\citenamefont#1{#1}\fi
\expandafter\ifx\csname url\endcsname\relax
  \def\url#1{\texttt{#1}}\fi
\expandafter\ifx\csname urlprefix\endcsname\relax\def\urlprefix{URL }\fi
\providecommand{\bibinfo}[2]{#2}
\providecommand{\eprint}[2][]{\url{#2}}

\bibitem[{\citenamefont{Langer}(1964)}]{Langer1964}
\bibinfo{author}{\bibfnamefont{W.~L.} \bibnamefont{Langer}},
  \bibinfo{journal}{Scientific American} pp. \bibinfo{pages}{114--121}
  (\bibinfo{year}{1964}).

\bibitem[{\citenamefont{Rvachev and Longini}(1995)}]{Rvachev95}
\bibinfo{author}{\bibfnamefont{L.~A.} \bibnamefont{Rvachev}} \bibnamefont{and}
  \bibinfo{author}{\bibfnamefont{I.~M.} \bibnamefont{Longini}},
  \bibinfo{journal}{Mathematical Biosciences} \textbf{\bibinfo{volume}{75}},
  \bibinfo{pages}{3} (\bibinfo{year}{1995}).

\bibitem[{\citenamefont{Murray}(2003)}]{Murray2003b}
\bibinfo{author}{\bibfnamefont{J.~D.} \bibnamefont{Murray}},
  \emph{\bibinfo{title}{Mathematical Biology {II}: Spatial Models and
  Biomedical Applications}} (\bibinfo{publisher}{Springer Verlag},
  \bibinfo{address}{New York}, \bibinfo{year}{2003}).

\bibitem[{\citenamefont{Durret and Levin}(1994)}]{durret94}
\bibinfo{author}{\bibfnamefont{R.}~\bibnamefont{Durret}} \bibnamefont{and}
  \bibinfo{author}{\bibfnamefont{S.}~\bibnamefont{Levin}},
  \bibinfo{journal}{Theoretical Population Biology}
  \textbf{\bibinfo{volume}{46}}, \bibinfo{pages}{363} (\bibinfo{year}{1994}).

\bibitem[{\citenamefont{Fuk{\'s} and Lawniczak}(2001)}]{paper15}
\bibinfo{author}{\bibfnamefont{H.}~\bibnamefont{Fuk{\'s}}} \bibnamefont{and}
  \bibinfo{author}{\bibfnamefont{A.~T.} \bibnamefont{Lawniczak}},
  \bibinfo{journal}{Discrete Dynamics Dynamics in Nature and Society}
  \textbf{\bibinfo{volume}{6}}, \bibinfo{pages}{191} (\bibinfo{year}{2001}),
  \eprint{arXiv:nlin.CG/0207048}.

\bibitem[{\citenamefont{{Di Stefano} et~al.}(2000)\citenamefont{{Di Stefano},
  Fuk{\'s}, and Lawniczak}}]{paper16}
\bibinfo{author}{\bibfnamefont{B.}~\bibnamefont{{Di Stefano}}},
  \bibinfo{author}{\bibfnamefont{H.}~\bibnamefont{Fuk{\'s}}}, \bibnamefont{and}
  \bibinfo{author}{\bibfnamefont{A.~T.} \bibnamefont{Lawniczak}}, in
  \emph{\bibinfo{booktitle}{Proceedings of Canadian Conference on Electrical
  and Computer Engineering, Halifax, May 2000}} (\bibinfo{year}{2000}), pp.
  \bibinfo{pages}{26--31}.

\bibitem[{\citenamefont{Benyoussef et~al.}(1999)\citenamefont{Benyoussef,
  Boccara, Chakib, and Ez-Zahraouy}}]{benyo2000}
\bibinfo{author}{\bibfnamefont{A.}~\bibnamefont{Benyoussef}},
  \bibinfo{author}{\bibfnamefont{N.}~\bibnamefont{Boccara}},
  \bibinfo{author}{\bibfnamefont{H.}~\bibnamefont{Chakib}}, \bibnamefont{and}
  \bibinfo{author}{\bibfnamefont{H.}~\bibnamefont{Ez-Zahraouy}},
  \bibinfo{journal}{Int. J. Mod. Phys. C} \textbf{\bibinfo{volume}{10}},
  \bibinfo{pages}{1025} (\bibinfo{year}{1999}).

\bibitem[{\citenamefont{Fuk{\'s} et~al.}(2005)\citenamefont{Fuk{\'s}, Duchesne,
  and Lawniczak}}]{paper24}
\bibinfo{author}{\bibfnamefont{H.}~\bibnamefont{Fuk{\'s}}},
  \bibinfo{author}{\bibfnamefont{R.}~\bibnamefont{Duchesne}}, \bibnamefont{and}
  \bibinfo{author}{\bibfnamefont{A.}~\bibnamefont{Lawniczak}}, in
  \emph{\bibinfo{booktitle}{Proceeding of 7th WSEAS International Conference on
  Applied Mathematics, Canun, Mexico, May 11-14 2005}} (\bibinfo{year}{2005}),
  pp. \bibinfo{pages}{108--113}, \eprint{arXiv:nlin.CG/0505044}.

\bibitem[{\citenamefont{Watts}(2004)}]{Watts2004}
\bibinfo{author}{\bibfnamefont{D.~J.} \bibnamefont{Watts}},
  \emph{\bibinfo{title}{Six Degrees: The Science of a Connected Age}}
  (\bibinfo{publisher}{W. W. Norton}, \bibinfo{year}{2004}).

\bibitem[{\citenamefont{Barrett et~al.}(2005)\citenamefont{Barrett, Eubank, and
  Smith}}]{Barrett2005}
\bibinfo{author}{\bibfnamefont{C.~L.} \bibnamefont{Barrett}},
  \bibinfo{author}{\bibfnamefont{S.~G.} \bibnamefont{Eubank}},
  \bibnamefont{and} \bibinfo{author}{\bibfnamefont{J.~P.} \bibnamefont{Smith}},
  \bibinfo{journal}{Scientific American} \textbf{\bibinfo{volume}{292}},
  \bibinfo{pages}{54} (\bibinfo{year}{2005}).

\bibitem[{\citenamefont{{Statistics Canada}}(2001{\natexlab{a}})}]{statcan1}
\bibinfo{author}{\bibnamefont{{Statistics Canada}}},
  \emph{\bibinfo{title}{Dissemination area digital cartographic file}},
  \bibinfo{howpublished}{Statistics Canada, Geography Division, Ottawa, ON}
  (\bibinfo{year}{2001}{\natexlab{a}}).

\bibitem[{\citenamefont{{Statistics Canada}}(2001{\natexlab{b}})}]{statcan2}
\bibinfo{author}{\bibnamefont{{Statistics Canada}}},
  \emph{\bibinfo{title}{Profile of age and sex, for {C}anada, provinces,
  territories, census divisions, census subdivisions, and dissemination areas,
  2001 census}}, \bibinfo{howpublished}{Industry Canada, Ottawa, ON}
  (\bibinfo{year}{2001}{\natexlab{b}}).

\bibitem[{\citenamefont{Sch{\"o}nfisch}(1993)}]{schon93}
\bibinfo{author}{\bibfnamefont{B.}~\bibnamefont{Sch{\"o}nfisch}}, Ph.D. thesis,
  \bibinfo{school}{Universit{\"a}t T{\"u}bingen} (\bibinfo{year}{1993}).

\bibitem[{\citenamefont{Boccara and Cheong}(1993)}]{bc93}
\bibinfo{author}{\bibfnamefont{N.}~\bibnamefont{Boccara}} \bibnamefont{and}
  \bibinfo{author}{\bibfnamefont{K.}~\bibnamefont{Cheong}},
  \bibinfo{journal}{J.Phys. A: Math. Gen.} \textbf{\bibinfo{volume}{26}},
  \bibinfo{pages}{3707} (\bibinfo{year}{1993}).

\bibitem[{\citenamefont{Duryea et~al.}(1999)\citenamefont{Duryea, Caraco,
  Gardner, Maniatty, and Szymanski}}]{duryea99}
\bibinfo{author}{\bibfnamefont{M.}~\bibnamefont{Duryea}},
  \bibinfo{author}{\bibfnamefont{T.}~\bibnamefont{Caraco}},
  \bibinfo{author}{\bibfnamefont{G.}~\bibnamefont{Gardner}},
  \bibinfo{author}{\bibfnamefont{W.}~\bibnamefont{Maniatty}}, \bibnamefont{and}
  \bibinfo{author}{\bibfnamefont{B.~K.} \bibnamefont{Szymanski}},
  \bibinfo{journal}{Physica D} \textbf{\bibinfo{volume}{132}},
  \bibinfo{pages}{511} (\bibinfo{year}{1999}).

\bibitem[{\citenamefont{Bonabeau et~al.}(1998)\citenamefont{Bonabeau, Toubiana,
  and Flahault}}]{Bonabeau1998}
\bibinfo{author}{\bibfnamefont{E.}~\bibnamefont{Bonabeau}},
  \bibinfo{author}{\bibfnamefont{L.}~\bibnamefont{Toubiana}}, \bibnamefont{and}
  \bibinfo{author}{\bibfnamefont{A.}~\bibnamefont{Flahault}},
  \bibinfo{journal}{J. Phys. A: Math. Gen.} \textbf{\bibinfo{volume}{31}},
  \bibinfo{pages}{L361} (\bibinfo{year}{1998}).

\end{thebibliography}
\end{document}